\documentclass[prl,twocolumn,superscriptaddress,floats,citeautoscript,showpacs]{revtex4}

\usepackage{graphicx}
\usepackage{amsmath}
\usepackage{amssymb}
\usepackage{times}
\usepackage{colordvi}
\usepackage{bm}

\newcommand* {\ee}{\ensuremath{\mathrm{e}}}
\newcommand{\ket}[1]{\left |  #1 \right \rangle}
\newcommand{\bra}[1]{\left \langle #1  \right |}

\graphicspath{{.}{./EPS/}}

\begin{document}

\title{Rotational Fluxons of Bose-Einstein Condensates in Coplanar Double-Ring Traps}

\author{J. Brand}
\affiliation{Centre for Theoretical Chemistry and Physics, Massey University (Albany Campus),
Private Bag 102~904, North Shore MSC, Auckland, New Zealand}
\affiliation{Institute of Fundamental Sciences, Massey University (Manawatu Campus),
Private Bag 11~222, Palmerston North, New Zealand}

\author{T.~J. Haigh}
\affiliation{Institute of Fundamental Sciences, Massey University (Manawatu Campus),
Private Bag 11~222, Palmerston North, New Zealand}

\author{U. Z\"ulicke}
\affiliation{Institute of Fundamental Sciences, Massey University (Manawatu Campus),
Private Bag 11~222, Palmerston North, New Zealand}
\affiliation{Centre for Theoretical Chemistry and Physics, Massey University (Albany Campus),
Private Bag 102~904, North Shore MSC, Auckland, New Zealand}

\date{February 26, 2009}

\begin{abstract}
Rotational analogs to magnetic fluxons in conventional Josephson junctions are predicted to emerge
{\em in the ground state\/} of 
rotating tunnel-coupled annular Bose-Einstein condensates (BECs).
Such topological condensate-phase
structures can be manipulated by external potentials. We determine conditions for observing
macroscopic quantum tunneling of a fluxon. 
Rotational fluxons in double-ring BECs can be created, manipulated, and controlled by external potential in different ways than possible in the solid state system, thus rendering them
a promising new candidate system for studying and utilizing quantum properties of collective
many-particle degrees of freedom.
\end{abstract}


\maketitle

Remarkable experimental advances have made it possible to engineer cold atom systems
to represent landmark models from completely different fields of physics. Examples include
quantum phase transitions~\cite{Bloch05} and the Josephson effect~\cite{albiez:prl:05}.
Besides intriguing nonlinear dynamics, the Josephson effect shows macroscopic quantum
phenomena with exciting prospects for applications~\cite{barone}. Long Josephson
junctions were used, e.g., to trap and study magnetic flux quanta, and the macroscopic
quantum tunneling of such fluxons was observed~\cite{wall:nat:03}. Here we report the
existence of topological condensate-phase structure equivalent to superconducting fluxons
in rotating BECs that are confined in
two concentric ring-shaped traps. The BECs in the individual rings are coupled by tunneling through a potential barrier.
The rotational fluxons 
can be understood as vortices that have entered the tunnel barrier. They
 show intriguing dynamical behavior and macroscopic quantum
properties. Easier accessibility and more straightforward means of manipulation 
than possible in conventional Josephson junctions
make
rotational fluxons in tunnel-coupled BECs attractive for investigating fundamental problems ranging from
models for cosmological evolution~\cite{KiZuJJ} to possibilities for realizing quantum
information processing~\cite{noriRev}.
 
\begin{figure}[b]
\includegraphics[width=2.8in]{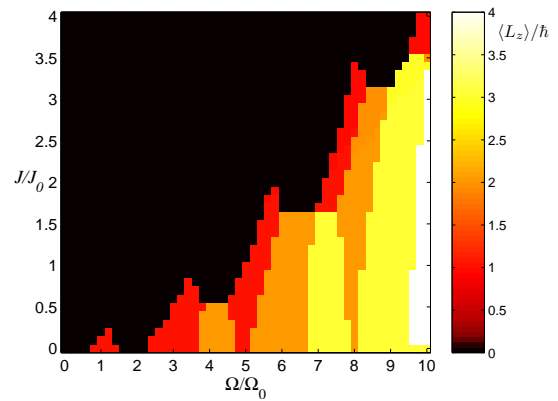}
\caption{\label{fig:mDiff}
(Color online) Phase diagram of co-planar double-ring BECs. We plot the difference
of angular-momentum expectation values for condensate atoms in the outer and inner
rings as a function of rotation frequency $\Omega$ and tunnel coupling $J$. Finite
integer values observed at higher frequencies are associated with the presence of
rotational fluxons. Results shown are obtained for a typical double-ring geometry
with $d=0.36$ and $g/J_0 = 100$.
Parameters and units are defined in the text.}
\end{figure}
\begin{figure}[t]
\includegraphics[width=2.8in]{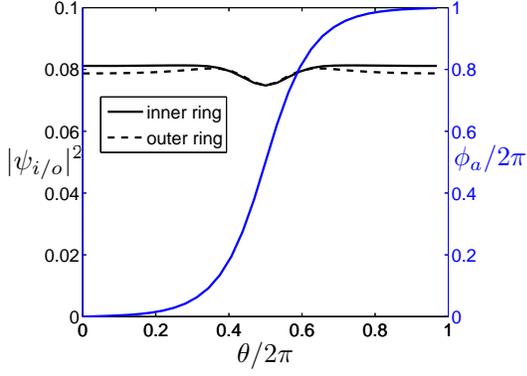}
\caption{\label{fig:singleJV}
(Color online) Single fluxon in the ground state of a double-ring BEC, signified by the
step-wise spatial variation of the relative phase $\phi_{\text{a}}$  for condensate
fractions in the two rings. Also shown are partial condensate densities
$|\psi_{\text{i/o}}|^2$ for the inner/outer ring. ($\theta$ is the azimuth.
Parameters are $\Omega=5.8 \Omega_0$, $J=1.9 J_0$, and those used in
Fig.~\ref{fig:mDiff}.)}
\end{figure}

Recent successful efforts to create annular trapping geometries~\cite{gupta:prl:05,Fernholz} 
and the routine use of trap rotation to simulate the effect a magnetic field has on charged 
particles~\cite{fetter:08} have motivated our present theoretical work. Unlike in the previously
considered cases of vertically separated double-ring traps~\cite{vKlitz:prl:07} or coupled
elongated BECs~\cite{bou:epjd:05,kau:pra:06}, 
we predict rotational fluxons to occur in the 
ground states of the proposed system.
In contrast to unconfined vortices in harmonically trapped BECs, rotational fluxons are confined to the tunnel barrier region between the coupled rings  for energetic reasons and thus
take on properties of {\em topological solitons} \cite{note}.
Preparing ground state solitons by cooling opens unprecedented opportunities for precision experiments on classical and quantum soliton dynamics. 
The phase structures are analogous
to magnetic flux quanta occurring in a superconducting Josephson junction in a parallel magnetic
field~\cite{barone}. 
In co-planar double-ring BECs, such fluxons appear due to a competition between trap
rotation and coherent tunneling.
While the former demands a tangential velocity $\Omega R$ different for the two rings,
the latter favors 
identical tangential velocities.
Spatially nonuniform or time-dependent potential differences between the two rings
as caused by gradient or curved magnetic, gravitational, or optical fields
generate forces on the fluxons. We also consider the conditions under which quantum tunneling
of fluxons may be observed. Further theoretical and experimental study of rotational fluxons in
BECs promises to shed new light on the behavior of collective excitations in interacting
quasi-onedimensional systems~\cite{gritsev} and their macroscopic quantum
properties~\cite{wall:nat:03}.

Below we start by presenting the theoretical description of a tunnel-coupled co-planar
double-ring system, which is based on the Gross-Pitaevskii (mean-field) equation with
a radial double-well potential. Its solution provides the ground-state phase diagram as
a function of trap-rotation frequency and tunnel-coupling strength, as shown in
Fig.~\ref{fig:mDiff}. For finite tunnel coupling and a slow rotation, the phase difference
between the two partial condensates in the individual rings vanishes. However,
beyond a critical value of the rotation frequency, a quantized relative-phase winding
between the two rings is accommodated, corresponding to a single rotational fluxon
whose phase and density profiles are illustrated in Fig.~\ref{fig:singleJV}. 
The phase structure can be detected experimentally by interferometry. E.g.\ after switching off the double-ring trap, both BECs will overlap in expansion and interfere destructively (resulting in a density node) at the azimuthal position of the fluxon. 
At higher
critical values of rotation frequency, the number of rotational fluxons successively
increases and a one-dimensional fluxon lattice is formed. After explaining our 
microscopic model and discussing numerical results, we present results from an effective 
hydrodynamic theory for BEC phase and density variables that captures the
numerically observed behavior. A classical equation of motion for rotational fluxons can
be derived, showing that time-dependent and/or spatially nonuniform potentials
accelerate fluxons. We find an expression for the fluxon's inertial mass and discuss
the possibility of quantum effects exhibited by these collective degrees of freedom.

\textit{Theoretical description of co-planar double-ring BECs\/}.
We consider a situation where a BEC of atoms having mass $M$ is confined by an
external magnetic and/or optical trapping potential to two concentric rings with, in 
general, different radii $R_{\text{i}}$ and $R_{\text{o}}$ for the inner and outer ring, 
respectively. We assume that transverse excitations in the individual rings are frozen out,
allowing for a purely one-dimensional description. In addition, the two rings are coupled
linearly by tunneling through a barrier with an associated tunnel energy $J$. Following 
Ref.~\onlinecite{vKlitz:prl:07}, we consider the coupled Gross-Pitaevskii equations for
the inner and outer ring wave functions $\psi_{\text{i}}(\theta,t)$ and $\psi_{\text{o}}
(\theta,t)$, respectively, which are given by
\begin{widetext}
\begin{equation} \label{iniGPE}
i \hbar \partial_t \psi_{\text{i/o}} =  \left[ -\frac{\hbar^2}{2 M R_{\text{i/o}}^2} \,
\partial^2_\theta + i \hbar \Omega \partial_\theta 
+\beta
\mp \delta + g_{\text{i/o}}
|\psi_{\text{i/o}}|^2 \right] \psi_{\text{i/o}} - J\, \psi_{\text{o/i}} \quad .
\end{equation}
\end{widetext}
Here $\delta = (E_{\text{o}}-E_{\text{i}})/2$ 
and $\beta = (E_{\text{o}}+E_{\text{i}})/2$ 
in terms of the single-well bound-state
energies $E_{\text{i/o}}$.
Rotation around the trap axis with frequency $\Omega$ is imposed by any (initial) anisotropy in the trapping potential.
 Using the the normalization condition $\sum_{\alpha=\text{i,o}}
\int |\psi_\alpha|^2 d\theta = 1$ the nonlinear coupling energies are $g_{\text{i/o}} = n
g_{\text{1D}}^{(\text{i/o})}$, where $n=N/(2 \pi R)$ is an average linear particle density
and $g_{\text{1D}}^{(\text{i/o})}$ is the effective one-dimensional coupling
strength~\cite{Olshanii1998a}.
For convenience, we introduce the effective trap radius $R=\sqrt{2} R_{\text{o}} R_{\text{i}}
/\sqrt{R_{\text{o}}^2+R_{\text{i}}^2}$ and $d=(R_{\text{o}}^2-R_{\text{i}}^2)/ (R_{\text{o}}^2
+R_{\text{i}}^2)$, which is a measure of the radial wells' separation, as parameters instead
of $R_{\text{i/o}}$. We have solved Eq.~(\ref{iniGPE}) using an FFT-based pseudospectral
method with imaginary-time propagation~\cite{imTimeProp} to find the ground states of 
double-ring BECs. For simplicity, we assumed $g_{\text{i}} = g_{\text{o}}\equiv g$. To 
compensate a trivial energy shift between states in the inner and outer well due to finite 
rotation, we 
have set $\delta$ to 
$\delta_\ast\equiv M \Omega^2 R^2 d/[2(1-d^2)]$ for Figs.~1 and 2.

In the absence of interactions (i.e., $g=0$), stationary solutions of Eq.~(\ref{iniGPE})
can be labelled by the quantum number $\hbar m$ of the angular-momentum component 
$L_z \equiv - i \hbar\partial_\theta$ along the symmetry axis of the trap.
The condensate wave functions in the inner/outer ring will be given by $\psi_{\text{i/o}}
(\theta) \propto e^{i m \theta}$, and the phase difference between condensate
amplitudes in the two rings will vanish at every point $\theta$. However, a finite $g$
introduces a mixing of amplitudes with different $m$ values in the condensate wave
function, enabling the appearance of nontrivial structure in the relative phase. To illustrate
this point quantitatively, we calculated the difference of expectation values of $L_z$
per particle in the outer and inner-ring condensate fractions, i.e., $\langle \Delta L_z\rangle
\equiv \langle L_z \rangle_{\text{o}} - \langle L_z \rangle_{\text{i}}$, where $\langle L_z
\rangle_{\text{i/o}}=\bra{\psi_{\text{i/o}}} L_z \ket{\psi_{\text{i/o}}}/\langle \psi_{\text{i/o}}
| \psi_{\text{i/o}}\rangle$. In Fig.~\ref{fig:mDiff}, $\langle \Delta L_z\rangle/\hbar$ is 
plotted as a function of tunnel coupling $J$ [measured in units of $J_0 = \hbar^2 / (2 M
R^2)$] and rotation frequency $\Omega$ [measured in units of $\Omega_0 = \hbar/ (2 M
R^2)$], for a particular double-ring geometry.
Regions with finite {\em integer\/} $\langle\Delta L_z \rangle/\hbar$ are observed, which
correspond to ground states with (one or more) rotational fluxons present. A
representative example for such a fluxon's relative-phase and partial-condensate
density profiles is shown in Fig.~\ref{fig:singleJV}.

Basic features of the phase diagram shown in Fig.~\ref{fig:mDiff} can be understood by a
variational consideration that assumes (i)~strong nonlinear coupling $g$ such that both
rings are populated with equal density, and (ii)~the condensate wave function in each ring
to be given by an $L_z$ eigenstate, $\psi^{(\mathrm{var})}_{\text{i/o}}(\theta) = e^{i
m_{\text{i/o}} \theta}/\sqrt{4\pi}$. The values of $m_{\text{i/o}} $ are determined by a
competition between tunneling, which tends to enforce equal phase for condensate 
fractions in both the inner and outer ring ($m_{\text{i}}=m_{\text{o}} \equiv m_\ast$), and
rotation. The latter favors the two condensate fractions to have, in general, different
angular momenta determined by the rotation frequency and the ring radii ($m_{\text{i/o}}=
\tilde m_{\text{i/o}}\equiv {\mathrm{Int}}[M R_{\text{i/o}}^2\Omega/\hbar +1/2]$). It is
straightforward to derive the energy functional of the system,
\begin{equation}
{\mathcal E}[m_{\text{o}},m_{\text{i}}] = \frac{\hbar^2}{4 M} \left( \frac{m_{\text{o}}^2}
{R_{\text{o}}^2} + \frac{m_{\text{i}}^2}{R_{\text{i}}^2} \right) - \frac{\hbar\Omega}{2}
\left(m_{\text{o}} + m_{\text{i}} \right) - J\, \delta_{m_{\text{o}}, m_{\text{i}}} .
\end{equation}
The condition ${\mathcal E}[\tilde m_{\text{o}}, \tilde m_{\text{i}}] = {\mathcal E}[m_\ast,
m_\ast]$ defines a critical value $J_{\text{cr}}\equiv M \Omega^2 R^2 d^2 /[2(1-d^2)]$.
For $J>J_{\text{cr}}$, the state having $m_{\text{i}}=m_{\text{o}}\equiv m_\ast$ would be
expected to be the ground state, corresponding to the black region in Fig.~\ref{fig:mDiff}.
In the opposite case, the phase gradient for partial-condensate wave functions in the two
rings will be different, essentially realizing a vortex (or several vortices) in the phase
difference between the two rings. Such a situation is signified by the brighter colored
regions in Fig.~\ref{fig:mDiff}. The variational estimate of $J_{\text{cr}}$ yields
a reasonably accurate description of the actual phase boundaries seen in the numerically
obtained phase diagram.

\textit{Effective analytical theory of fluxon phase profile and dynamics\/}.
To obtain a more detailed understanding of fluxons in coupled annular BECs, we
consider the dynamics of their collective phase and density variables. This approach
applies equally well to co-planar and vertically separated double-ring traps. Writing the
partial-condensate wave functions as $\psi_{\text{i/o}} = |\psi_{\text{i/o}}|\exp
\{i\phi_{\text{i/o}}\}$, we define symmetric and antisymmetric combinations of their
modulus and phase and express the Lagrangian of the double-ring system in terms of
these new quantities. It is possible to derive a closed equation of motion for the phase
difference $\phi_{\text{a}} = \phi_{\text{o}} - \phi_{\text{i}}$ that is accurate to first order in
the typically small quantity $E_R/g$, where $E_R = \hbar^2/(2M R^2)$ is the scale of
energy quantization on the ring. Its lengthy analytical expression is omitted.

In the stationary limit and to leading (zeroth) order in $E_R/g$, we find
\begin{equation}
(1-d^2) E_R \, \partial_\theta^2 \phi_{\text{a}} - 2 J \,  \sin\phi_{\text{a}} =0
\quad ,
\end{equation}
which has a single-soliton (i.e., fluxon) solution~\cite{ziv:prb:94}
\begin{equation}
\phi_{\text{a}}^{\text{(fl)}} (\theta, \theta_0) = \pi+ 2 \, {\mathrm{arcsin}} \left[ {\mathrm{sn}}
\left( \left. \frac{\kappa (\theta - \theta_0)}{k} \right| k \right) \right] .
\end{equation}
Here $\mathrm{sn}(u|k)$ is a Jacobi elliptic function~\cite{abramowitz} whose
parameter $k$ is determined from the transcendental relation
\begin{equation}\label{eq:kDeterm}
\pi \kappa = k K(k) \quad ,
\end{equation}
involving the complete elliptic integral of the first kind, and $\kappa=\sqrt{2 J/[(1-d^2)
E_R]}$. Hence, fluxons emerge as stationary phase configurations, as seen in our
numerical calculations. The dimensionless parameter $\kappa \equiv R /(\sqrt{1-d^2}
\lambda_J)$ can be interpreted as the ratio of the quadratic mean radius of the trap
$R /\sqrt{1-d^2} = \sqrt{(R_o^2 + R_i^2)/2}$ and the physical length scale of the fluxon
$\lambda_J = \hbar/(2 \sqrt{MJ})$, which is set by the tunnel coupling.

To obtain a dynamical equation for a slowly moving fluxon, we insert the ansatz
$\phi_{\text{a}}(\theta, \tau)=\phi_{\text{a}}^{\text{(fl)}}(\theta, \theta_0(t))$ into 
the equation of motion for the phase difference. Here $\theta_0(t)$ is the instantaneous
position of the fluxon. Straightforward algebraic manipulation yields a Newton-like
equation of motion:
\begin{equation} \label{eq:neom}
M_{\text{fl}} \ddot{\theta}_0 R =  {F}_{\text{fl}}\quad.
\end{equation}
The fluxon's dynamical mass is $M_{\text{fl}} = 2 \sqrt{E_R/g} \, {\mathcal I}_{\text{fl}} M$
with the dimensionless moment of inertia given by
\begin{subequations}
\begin{eqnarray} \label{eq:JVinertia}
{\mathcal I}_{\text{fl}} &=& (1+d^2) \int_0^{2\pi} \!\! \frac{d\theta}{4\pi} \, \left[
\partial_\theta\phi_{\text{a}}^{\text{(fl)}}\right]^2 \quad , \\ \label{eq:JVinertiaA}
&\equiv& \frac{1+d^2}{\pi}\, \frac{\kappa}{k} \,\, {\mathrm{E}} \left(\left. \frac{2\pi\kappa}{k}
\right| k\right)\quad ,
\end{eqnarray}
\end{subequations}
where $E(u|k)$ is the incomplete elliptic integral of the second kind~\cite{abramowitz}. The
general expression for the force (torque) on the fluxon is~\cite{kappaForce}
\begin{widetext}
\begin{equation}\label{eq:JVtorque}
{F}_{\text{fl}} = \sqrt{\frac{2 M}{g}} \int_0^{2\pi} \frac{d\theta}{2\pi} \,\,\, \partial_\theta
\phi_{\text{a}}^{\text{(fl)}} \left\{ \partial_t \delta 
+ d \, \partial_t \beta
+ d \left( \frac{(1-d^2) E_R}{2\hbar}\,
\partial_\theta \phi_{\text{a}}^{\text{(fl)}} - \Omega d \right) \partial_\theta \delta \right\}
\quad .
\end{equation}
\end{widetext}

\begin{figure}[t]
\includegraphics[width=0.95\columnwidth]{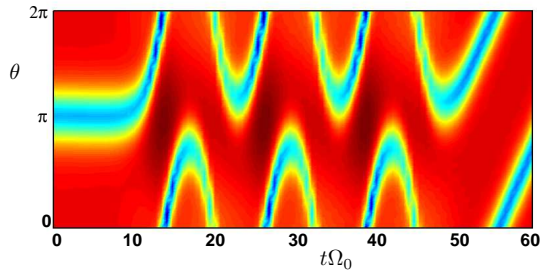}
\caption{\label{fig:JVinertia}
(Color online) Dynamics of a single fluxon under a time-dependent external field according to Eq.\ (\ref{iniGPE}) (color coded density $|\psi_i|^2 + \psi_o|^2$). An external potential gradient across the double-ring trap corresponding to 
$\delta = 0.2 \beta = 0.2 J_0 \cos(\theta)$ is smoothly turned on  at $t=10 \Omega_0^{-1}$ and off again at $t=50 \Omega_0^{-1}$. Other parameters are $J=J_0$ and $g=100 J_0$.}
\end{figure}

Equations (\ref{eq:JVinertiaA}) and (\ref{eq:kDeterm}) define a universal relationship between
the fluxon's dimensionless moment of inertia ${\mathcal I}_{\text{fl}}$ and the variable $\kappa$.
The limiting value of ${\mathcal I}_{\text{fl}}$ for small trap size
($\kappa \ll 1$) is a constant [$(1+d^2)/2$], whereas a linear dependence [$2 (1+d^2)
\kappa/\pi$] is realized for large ring traps ($\kappa \gg 1$). 

Inspection of Eq.~(\ref{eq:JVtorque}) reveals that fluxons subject to spatially nonuniform
$\delta$ and/or time-dependent $\delta$ or $\beta$ will experience a force. This feature is
confirmed by our numerical solution of Eq.~(\ref{iniGPE}), an example being shown in Fig.~\ref{fig:JVinertia}. In the case of spatially uniform $\delta(\theta, t) \equiv \delta_0(t)$
and $\beta= 0$,
the force simplifies to $\sqrt{2M/g}\, \sigma \dot\delta_0$, 
which is similar to the result found previously~\cite{kau:pra:06} for phase-imprinted fluxons
in a junction between two parallel {\em linear\/} BECs.
The sign $\sigma = {\mathrm{sgn}} \Big[ \phi_{\text{a}}^{\text{(fl)}}(2\pi) -
\phi_{\text{a}}^{\text{(fl)}}(0) \Big]$ is the topological charge of the fluxon
related to
its orientation. Here we found the expression for the force felt by fluxons in the more general
case with $d\ne 0$. 

If the external fields are time-independent and the fluxon length $\lambda_J$ is smaller
than the length scale of spatial variations of $\delta$, Eq.~(\ref{eq:JVtorque}) can be
integrated and written as $F_{{\text fl}} = - R^{-1} \partial_\theta V$, where $V$ is a
potential energy. For $\kappa \gg 1$ and to leading order in $d$, we obtain
\begin{equation}
  V (\theta) = \left(\sigma \frac{\hbar \Omega d^2}{\sqrt{E_R g}} - \frac{2\sqrt{2} d}{\pi}
  \sqrt{\frac{J}{g}} \right)\delta(\theta) ,
\end{equation}
for the potential and $M_{\text{fl}} = M \sqrt{32 J /(\pi^2 g)}$ for the dynamical fluxon mass.

\textit{Macroscopic quantum tunneling\/}. 
Describing the effects of quantum and thermal fluctuations on the fluxon dynamics can 
proceed in analogy to the established treatment of Josephson vortices in superconducting
junctions~\cite{barone}. In particular, the possibility of fluxon (macroscopic quantum)
tunneling can be included~\cite{kato:jpsj:96} by direct quantization of the classical equation
of motion (\ref{eq:neom}). A rough estimate for tunneling of a fluxon through a potential
barrier of height $\Delta V$ and length $\Delta l$ from the WKB method yields the probability
$P \approx \exp(-2 \Delta l \sqrt{2 M_{\text{fl}} \Delta V}/\hbar)$. In order to have $P \gtrsim
1/\ee$  with $\Delta l\approx \lambda_J$, we need $\sqrt{Jg} \gtrsim \Delta V$. Assuming a
double-ring configuration as proposed in Ref.~\cite{Fernholz} with $R\approx 50\,\mu$m, it
may be feasible to achieve $g/k_B \approx 2\, \mu$K and $J/k_B \approx 0.05\, \mu$K
and observe quantum tunneling through barriers $\Delta V/k_B \lesssim 0.3\, \mu$K at
sufficiently low temperatures.

Quasiparticle excitations present at finite temperature will act as a damping mechanism for
fluxon motion~\cite{barone} and, at the same time, as a  source of quantum decoherence
(thus suppressing fluxon tunneling~\cite{kato:jpsj:96}).

\textit{Discussion and conclusions\/}.
We have discovered fluxon-like topological structure in the relative phase of condensate
fractions in the ground state of BECs in rotating double-ring traps. These rotational
fluxons are accelerated by spatially varying external potentials that couple asymmetrically
to the two rings and/or time-dependent potentials. 
Macroscopic
quantum tunneling of fluxons 
may become observable and would serve the long sought goal of preparing macroscopic quantum superposition states of BECs (see e.g.\ Ref.~\cite{Kuritzki}).
Future studies will focus on details of
fluxons' quantum properties and possible applications~\cite{KiZuJJ,noriRev}.

JB is supported by the Marsden Fund Council (contract MAU0706) from Government
funding, administered by the Royal Society of New Zealand.


\begin{thebibliography}{20}
\bibitem{note}
  {In a regime of strong tunnel coupling, a transition to a (non-topological) dark soliton can occur \cite{kau:pra:06}.}
\bibitem{kappaForce}
We assumed a spatially uniform $J$. For Josephson vortices in superconductors~\cite
{kato:jpsj:96}, forces arising due to a spatially varying tunnel coupling $J$ have been
discussed.
\end{thebibliography}

\end{document}